\documentclass[10pt, conference, compsocconf]{IEEEtran}  
\usepackage{tabularx}
\usepackage{graphicx}
\usepackage{adjustbox}
\usepackage{microtype}
\usepackage{multirow}

\usepackage{amsmath}
\usepackage{amssymb}
\usepackage{amsfonts}

\usepackage{scalerel}
\usepackage{pifont}

\usepackage{varwidth}

\usepackage{mathtools}
\usepackage{mathptmx}
\usepackage{subfigure}
\usepackage{bbm}

\usepackage{array}
\usepackage{latexsym}
\usepackage{stmaryrd}
\usepackage{color}
\usepackage{epsfig}
\usepackage{boxedminipage}
\usepackage{times}
\usepackage{verbatim}
\usepackage{url}
\usepackage{xspace}

\usepackage{verbments} 
\usepackage{paralist}
\usepackage{gensymb} 
\usepackage{graphicx}
\usepackage{caption}
\usepackage{listings}
\usepackage{xcolor}

\colorlet{punct}{red!60!black}
\definecolor{background}{HTML}{EEEEEE}
\definecolor{delim}{RGB}{20,105,176}
\colorlet{numb}{magenta!60!black}
\usepackage{lipsum}

\begin{document}

\title{SiEVE: Semantically Encoded Video Analytics on Edge and Cloud}

\author{\IEEEauthorblockN{Tarek Elgamal\IEEEauthorrefmark{1},
Shu Shi\IEEEauthorrefmark{4}\IEEEauthorrefmark{5}\thanks{\IEEEauthorrefmark{5}Work was done when the author was previously working for AT\&T Labs}, Varun Gupta\IEEEauthorrefmark{3}, Rittwik Jana\IEEEauthorrefmark{2}, Klara Nahrstedt\IEEEauthorrefmark{1}}
\IEEEauthorblockA{ \IEEEauthorrefmark{1}University of Illinois Urbana-Champaign, \IEEEauthorrefmark{2}AT\&T Labs Research, \IEEEauthorrefmark{3}Facebook, Inc.,\IEEEauthorrefmark{4} ByteDance\\
Emails: telgama2@illinois.edu,
pkuslick@gmail.com,
varung.iitd@gmail.com,
rjana@research.att.com, klara@illinois.edu
}
}

\maketitle

\begin{abstract}
Recent advances in computer vision and neural networks have made it possible for more surveillance videos to be automatically searched and analyzed by algorithms rather than humans. This happened in parallel with advances in edge computing where videos are analyzed over hierarchical clusters that contain edge devices, close to the video source. However, the current video analysis pipeline has several disadvantages when dealing with such advances. For example, video encoders have been designed for a long time to please human viewers and be agnostic of the downstream analysis task (e.g., object detection). Moreover, most of the video analytics systems leverage 2-tier architecture where the encoded video is sent to either a remote cloud or a private edge server but does not efficiently leverage both of them. In response to these advances, we present SIEVE, a 3-tier video analytics system to reduce the latency and increase the throughput of analytics over video streams. In SIEVE, we present a novel technique to detect objects in compressed video streams. We refer to this technique as \emph{semantic video encoding} because it allows video encoders to be aware of the semantics of the downstream task (e.g., object detection). Our results show that by 
leveraging \emph{semantic video encoding}, we achieve close to 100\% object detection accuracy with decompressing only 3.5\% of the video frames which results in more than 100x speedup compared to classical approaches that decompress every video frame.

\end{abstract}

\section{Introduction}\label{sec:intro}
Cameras are ubiquitous as the reports by the Information Handling Services show; their reports indicate that 245 million professionally installed surveillance cameras are operating worldwide as of 2015~\cite{IHS}. Such cameras are used for several purposes including: traffic control, surveillance, and security in both public
and private venues. Analyzing live video streams from those cameras is of considerable importance for decision making in many organizations such as traffic, police, and private security departments.  

A common objective of such cameras is object detection and recognition in each video frame. Due to the limited computational capability of the camera devices, the conventional approach of performing object detection on camera streams is to send the streams to a centralized data center (cloud) and leverage its powerful and seemingly abundant resources to execute the analytics remotely. However, given the tremendous amount of data transfer required by video data, the latency and bandwidth requirements become significantly high (e.g., 300GB/month for Nest cameras~\cite{nest}). For this reason, the concept of cloudlet/edge computing~\cite{cloudlet} has emerged in which an additional computing layer sits between the camera and the cloud, and it is used to help reduce the bandwidth and latency by performing the entire computations on behalf of the cloud or performing simple computations to filter the amount of data being transmitted to the cloud. The additional compute devices known as edge devices/servers complement the cloud-centric approach resulting in a 3-tier architecture (i.e., cameras, edge devices, and cloud servers). 

Most of the existing object detection~\cite{noscope}~\cite{chameleon} systems for live video streams do not leverage the 3-tier architecture, and instead they focus on a 2-tier architecture where the encoded video is sent from the camera/smartphone to either a remote cloud server or a private edge server. The existing 2-tier approaches can reduce either transmission latency by utilizing an edge server, or computation latency by utilizing more powerful cloud servers, unlike 3-tier architectures which can make the appropriate tradeoff between optimizing both. 
However, utilizing 3-tier architecture poses several challenges to speed up object detection on video streams. A major challenge that we address in this paper is to detect if the current frame has different objects than the previous frames without decoding/decompressing the full video. Addressing this challenge can significantly reduce the amount of data sent from the edge to the cloud and avoids decoding and executing expensive object detection computation on every frame of the video. Existing approaches leverage cheap image similarity computation (e.g., mean squared error) to solve this problem. They avoid transmitting frames that do not show significant difference from their previous counterparts, however, this requires unnecessarily decoding the entire video which is also a computationally intensive task.


To address this challenge, we present SiEVE
\footnote{Sieve is a tool used for separating coarser from finer particles. In our system we separate frames that are likely to have objects from other frames}, a 3-tier video analytics system to reduce the latency of NN-based analysis over video streams. In SiEVE, we focus on object detection. Given a target video and a reference pre-trained object detection neural network (NN), SiEVE detects if a new object enters or leaves the scene without decompressing the full video. Then SiEVE uses the reference NN to detect the actual object that appears in the scene. Hence, SiEVE marginally reduces the accuracy of frame-by-frame object detection, but achieves a significant improvement in bandwidth and latency when objects do not change frequently.  In summary, we make the following technical contributions:


(1) We present a system for accelerating neural network analysis over video streams  across geo-distributed resources. The system allows: (a) automatically detecting frames of interest in a compressed video, (b) deploying/partitioning NN layers across geo-distributed resources.


(2) A novel technique to tune video encoders to detect absence/presence of a particular object class in a video, and compress the video accordingly. This technique then alleviates the need to decompress the video for detecting absence/presence of such objects during subsequent analysis on edge/cloud devices.


(3) A complete end-to-end implementation and extensive evaluation of the system using a diverse set of videos that vary in the type of objects, resolution, and the geo-location in which the video was captured. Our results show that for a dataset of 2.16 million video frames, SiEVE can process 10 times more frames per second (fps) compared to image-similarity-based solutions, with 1-5\% loss in accuracy compared to the ground truth object labels. Moreover, SiEVE can reduce the size of the data transmitted to the cloud by a factor of 7x compared to the original compressed videos (12.26GB-1.68GB).

The rest of the paper is organized as follows. In Section~\ref{sec:related}, we discuss the related work. In Section~\ref{sec:system}, we present an overview of the system environment and the system architecture. Section~\ref{sec:encoder} describes the details of our semantic video encoder. 
We experimentally evaluate the system in Section~\ref{sec:eval}. Section~\ref{sec:conc} concludes the paper.

\section{Related Work}\label{sec:related}

Several techniques have been proposed to optimize the latency, bandwidth and accuracy of NN inference on videos. The techniques range from : (1) {\bf Video aspects}: selecting the best resolution, frame rate, and bitrate~\cite{chameleon}, to (2) {\bf NN aspects}: NN compression, pruning, fusion, lower precision, and specialized models~\cite{noscope}, all the way to (3) {\bf Hardware aspects}: Google's tensor processing unit~\cite{tpu}.

Putting our contribution in perspective, we focus on a video aspect ({\it Semantic Video Encoding}) that was not explored extensively in the related work. In {\it semantic video encoding},  we make video compression algorithms aware of the downstream analysis to improve the bandwidth, latency, and  reduce the amount of decompressed video frames. NoScope~\cite{noscope} tries to achieve the same goal of reducing the number of frames undergoing NN inference however their approach depends on computing image similarity between consecutive frames (e.g., SIFT matching~\cite{sift}, and mean squared error (MSE)). The fundamental novel aspect in our approach is leveraging the motion estimation in video encoders to generate I-frames when a significant motion difference exists. Contrary to existing methods, our method alleviates the need to decode the huge number of P-frames ($\approx$96\% of the video) which results in significantly faster analysis. {\it To the best of our knowledge, this is the first work to propose tuning video encoders to detect changes in object labels across a video.}

{\bf Tuning video parameters for NN Inference:} Several systems have proposed to tune the video parameters such as resolution, bitrate, and frame rate to achieve the same analysis accuracy\cite{chameleon} at an increased speed or lower bandwidth consumption. However, none of these systems explore frame rate below 1 frame per second. Unlike these approaches, our system can filter out a significant number of frames when the events change infrequently (e.g., once every few minutes). Moreover, our system takes in consideration the motion happening between two frames (i.e., scenecut threshold) which is more robust than sampling uniformly every 30 frames. Our system can, however, benefit from these approaches in tuning the parameters that we do not address (e.g., bitrate).

\section{System Overview}\label{sec:system}

\begin{figure}[t]
  	\centering
  	\includegraphics[width=\linewidth]{figures/sieve_storage_outside.pdf}
	\vspace{-0.2in}
	\caption{Proposed System Architecture.}
	\label{fig:overview}
	\vspace{-0.2in}
\end{figure}

Figure~\ref{fig:overview} shows the 3-tier architecture of the proposed video analytics system. The dashed lines in Figure~\ref{fig:overview} show the control commands and the solid lines show the data flow. The control commands are sent by a surveillance operator who is a dedicated personnel hired by an organization (e.g., the traffic department) to control the cameras and monitor the activity happening in the system. The operator can control the parameters of the video encoder such as \emph{GOP (Group of pictures) size}, and \emph{scenecut threshold}. We refer to the video encoder with controllable parameters as the {\emph Semantic Video Encoder}. The parameters of the semantic encoder are configured offline for each camera. We provide more details about the parameters and the techniques used to tune them in Section~\ref{sec:encoder}. The semantic encoder is designed to produce an I-frame (key frame) when it is more likely that this frame has objects that are different from the previous frame (e.g., a new car entered or left the scene). On the other hand, the non I-frames are likely to have the same object labels as the previous I-frame so they do not need to be analyzed separately but they get stored in the edge storage for further analysis beyond object detection (e.g., object tracking, person identification).

The edge server receives the {\it semantically encoded} video from the camera via a secure protocol that the camera supports (e.g., https or rtmps). The edge server then passes the {\it semantically encoded} video to an {\it I-frame seeker} module in which only the I-frames are extracted to be processed by the downstream neural network to identify if a new object entered or an existing object left the scene. However, the non I-frames (i.e., P-frames) will not be processed by the downstream neural network and they are assigned the same object labels as the previous I-frame. 
We note that the I-frame seeker is not actually decoding each frame in the video but instead it searches through the video metadata and {\it drops} every frame that is not of type I-frame. Our empirical results show that such I-frames are no more than 3.5\% of the entire video. Hence, our system saves a huge amount of computation load that is performed in the regular video decoding pipeline such as bit stream decoding, motion compensation, and Inverse Fourier Transform (IFT) for every frame. 



The frames that pass the I-frame seeker module are temporarily buffered in an event queue before being dispatched by the edge compute engine. The edge compute engine is a dataflow engine that takes an I-frame as an input from the event queue, decompresses it in the same way still JPEG images are decompressed, and passes the decompressed frame through multiple layers of the neural network (NN) model for object classification. The number of NN layers deployed on the edge compute engine is decided beforehand by the {\it NN Deployment service}. The deployment service can choose to: (1) deploy all NN layers in either the edge or the cloud compute engine, or (2) deploy a subset of the layers in the edge engine and the rest in the cloud engine. In this paper, we focus on the former, however, our system can leverage NN partitioning algorithms in the literature~\cite{neurosurgeon}. 
Based on the choice made by the NN deployment service, the edge compute engine computes the output of the sub-NN deployed in it and passes its output to the cloud compute engine over a secure http connection. The cloud engine computes the output of the rest of the neural networks layers deployed on it and stores the result in a database. The results are in the form of a list of tuples where each tuple consists of frame ID and the object names that appear in the frame.


\section{Semantic Video Encoder}\label{sec:encoder}
Video compression algorithms have been designed aiming at increasing the compression ratio, reducing the encoding/decoding time, and pleasing human viewers. However, since more and more of surveillance videos are going to be watched by algorithms, we propose an approach where we can train video compression algorithms to be aware of the downstream object detection task and produce key-frames only when a semantic event happens (e.g., new object enters the scene). In such cases, when a video is analyzed at real-time, there is no need to decompress each frame of the video. Only key-frames are seeked and decompressed independently the same way as still JPEG images are decompressed. The key-frame seeking and decompression are performed at the edge device located close to the camera. The decompressed I-frames are passed to the downstream NN to identify the new object that entered or left the scene. The NN computation can be performed in the edge device or in the cloud based on the required latency and the available compute resources on each device. In Section~\ref{sec:end-to-end}, we show the end-to-end system's performance when performing the NN in the edge and in the cloud.

{\bf Video Encoder Parameters:} To tune video encoders, we focus on two parameters: {\it scene cut threshold} and {\it GOP size}. We chose these two parameters because they control the number of I-frames and the duration between two I-frames. The \emph{scenecut threshold} is a threshold on the motion difference between two consecutive frames. It controls how aggressively I-frames need to be inserted. The higher the scenecut threshold value, the more sensitive it is to small motion and the more aggressive it places I-frames. Therefore, when the scenecut threshold is set to a high value (maximum 400) then more I-frames are created compared to setting the scenecut threshold to a low value (e.g., 20). On the other hand, the \emph{GOP size} is the duration between two I-frames (key frames). It is essentially the number of P and B frames between two I-frames. 



{\bf Offline Tuning:}  Due to the differences in camera positions and orientations, our approach focuses on tuning the encoder parameters (i.e., \emph{GOP size} \&  \emph{scenecut threshold}) for each camera independently. For example, we tune our parameters to find objects in the {\it "Jackson town square"} surveillance camera feed. To understand why we tune each camera separately, let us consider two cameras placed at the height of 5 and 10 meters from the road, respectively. The cars in the second camera will appear smaller (i.e., consume less number of pixels) than the cars in the first camera. Hence, the amount of motion (i.e., scenecut threshold) that signals a car entering the scene is smaller in the second camera compared to the first camera.

Our approach to tune the encoder parameters is to leverage historic labelled data that show the event of interest. For example, we collect several hours of video from a surveillance camera and we label events such as a new object entered the scene or an object that used to be in the scene is not visible any more. We can then tune the parameters of the video encoder based on the labelled events and we use the tuned parameter to detect future events in real-time.

To understand how we define events, we take an example of a 30 seconds video in which the scene has no cars for 10 seconds, then a car enters the scene and remains there for 10 seconds before it leaves the scene. We define 3 events in this video, where each event has 300 frames (10 seconds * 30 fps) and all frames within one event have the same object label. The first event has no label, the second event has the label car, and the third event has no label. We define the best event detection algorithm as the one that outputs the first frame of each event. This ensures that assigning the same object label for subsequent frames within the event will result in correct object labels.

Figure~\ref{fig:encoder} shows the detailed steps of tuning video codec parameters. We note that these steps are performed offline to find the best video encoding parameters for a given camera feed. The best parameters are then stored in a lookup table to be used for real-time event detection. The steps are described as follows:

\begin{figure}[t]
  	\centering
  	\includegraphics[width=\linewidth]{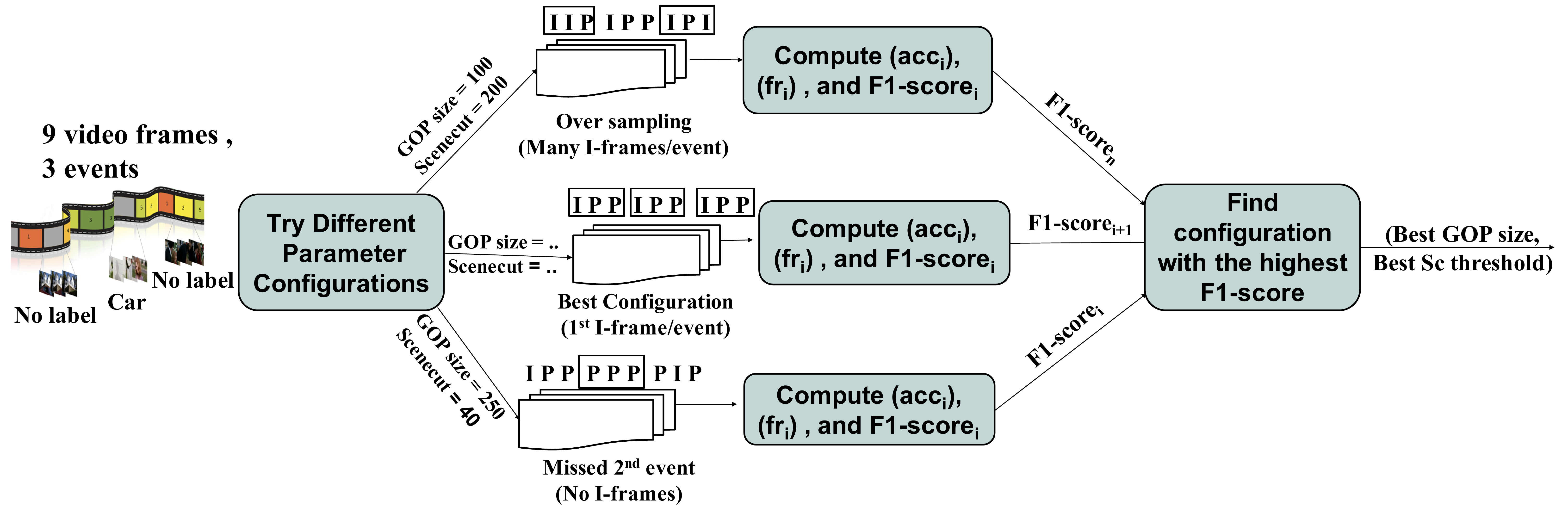}
	\caption{Steps of finding the best video encoder configuration for detecting object changes in compressed videos (Offline stage).}
	\label{fig:encoder}
    \vspace{-0.1in}
\end{figure}



(1) {\bf Step 1: } Instead of using the default parameters (i.e., GOP size = 250, and scenecut = 40), we try different configurations for the two parameters {\it GOP size} and {\it scenecut threshold} offline using historical data. We experiment with the $k$ values for GOP size (e.g., 100,250,1000,5000) and $l$ values for scenecut threshold (e.g., 20,40,100,200,250), so the total number of configurations is $k*l$. For each configuration, we re-encode the video with the corresponding parameter values. The result of this step is $k*l$ videos where each video has different numbers and positions of I-frames. The values of $k$ and $l$ define how many configurations are explored. The more configurations are explored, the more likely it is to get a better semantically encoded video (i.e., one I-frame per event). Since this process is done offline, the values of $k$ and $l$ are not critical to the real-time analysis of the videos. We choose five configurations for each parameter (i.e., $k=5$ and $l=5$). 
    
    
(2) {\bf Step 2: } We evaluate each parameter configuration $i$ (i.e., ($GOP_i$, $scenecut_i$)) by two metrics: (1) the accuracy of the event detection denoted by $acc_i$, and (2) the filtering rate, denoted by $fr_i$. The $acc_i$ is calculated based on the positions of I-frames in the encoded video. If each event starts with an I-frame, then the accuracy is 100\%. However, if an I-frame only appears in the middle of an event then the accuracy is reduced by the percentage of frames from the start of the event until this I-frame with respect to the total number of frames in the video.  
On the other hand, $fr_i$ is calculated as the ratio between the number of non I-frames and the total number of frames. We note that there is a tradeoff between the $acc_i$ and $fr_i$ because with more I-frames the $acc_i$ is likely to increase but the filtering rate decreases. To combine the two metrics in one quality metric, we calculate the harmonic mean (F1-score) between $acc_i$  and $fr_i$. The F1-score is measured as:
    
	$$F1score_i = \frac{2*acc_i*fr_i }{acc_i + fr_i}$$

    
(3) {\bf Step 3:} We choose the configuration that has the highest F1-score:$\quad i^* = \underset{i} {\mathrm{argmax}} \big(F1score_i\big)$

	The configuration with the highest F1-score balances the tradeoff between trying to filter as much redundant information as possible and getting a high event detection accuracy. 
	


{\bf Online Usage of Tuned Parameters:} 
The best encoding parameters for each camera are stored in a lookup table. The parameters are entered by the system's operator in the software provided by the camera's vendor as shown in Figure~\ref{fig:overview}. The new parameters will then be used for  real-time encoding of future live videos. The semantically {\it encoded} live video including I and P frames is then received at real-time by the I-frame seeker module (Figure~\ref{fig:overview}) that sits on the edge device which in turn searches for I-frames and sends them to the event queue for further processing by a neural network. 


{\bf Use cases:} The current prototype of the semantic encoder focuses on detecting the existence of new objects in surveillance cameras. It has its best results when the camera has a fixed-angle and the objects entering the scene create significant motion differences. We use that technique to detect the object labels in each frame without decompressing the majority of the frames. The object labels for each frame can then be used to do further analysis such as object tracking and person identification. The semantically encoded video that we store in the edge helps to quickly seek the exact event/GOP that can be further analyzed which significantly speeds up the analysis. A limitation in our approach is that we assume that the edge location has access to non-trivial storage capacity. We also note that several cameras have hardware encoders built into them with limited control over their parameters. In these cases, we re-encode the video with the semantic parameters on the edge device. 

\section{Evaluation}\label{sec:eval}

\noindent {\bf System setup:} The computing infrastructure consists of edge and cloud resources. We use one desktop as the edge device and one server as the cloud. The edge device has Intel Core i7-5600 CPU with 12 GB of memory and the cloud server has Intel Xeon E5-1603 CPU with 32 GB of memory. We control the bandwidth from edge to cloud server to be 30 Mbps which simulates an average wide area network connection. Each of the edge and cloud servers has a local dataflow engine, Apache NiFi, that handles execution of operators that are deployed on it. Nifi is an engine designed for composing user-defined operators and executing dataflows in a single machine or across multiple machines. Each of the edge and cloud servers has a local deployment of Nifi and we use Echo~\cite{echo} orchestration framework to handle the communication between the two Nifi instances.


{\bf Datasets:} We experiment with five different datasets. The five datasets vary in the object types (car, bus, truck, person, boat), resolutions (400p, 720p, 1080p), and locations (indoor, outdoor, different cities) as shown in Table~\ref{tab:datasets}. The first three datasets have publicly available ground truth object labels. We  experiment with 8 hours for each of them. The first 4 hours are used as a training set to tune the encoder parameters for our approach and the thresholds for the compared approaches. We use the next 4 hours for evaluation. The last two datasets are obtained from YouTube live feeds and they are used to evaluate the end-to-end system performance.

\begin{table*}
\centering

\begin{adjustbox}{width=0.8\textwidth}

\begin{tabular}{|c|c|c|c|c|c|c|}
\hline \bf Dataset name  & \bf Object & \bf Resolution & \bf FPS &\bf Duration &\bf Description &\bf labels ?     \\ \hline
Jackson square~\cite{noscope-github} & car, bus, truck & 600x400 & 30 & 8 hours & vehicles going back and forth in a public square  & Yes\\\hline
Coral reef~\cite{noscope-github} & person & 1280x720 & 30 & 8 hours & people watching coral reefs in an aquarium & Yes \\\hline
Venice~\cite{noscope-github}  & boat & 1920x1080 & 30 & 8 hours & boats moving in the lagoon & Yes \\\hline
Taipei~\cite{taipei}  & car, person & 1920x1080 & 30 & 4 hours & vehicles and people in a public square in Taipei & No \\\hline

Amsterdam~\cite{amsterdam}  & car, person & 1280x720 & 30 & 4 hours &  Road intersections in amsterdam & No \\\hline


\end{tabular}
\end{adjustbox}
\caption{ Datasets used in the evaluation. }
\label{tab:datasets}
\vspace{-0.1in}

\end{table*}




 In the following, we conduct experiments to evaluate the improvement of the event detection module separately. We then present results for the improvement in the end-to-end performance of the system.

\subsection{Evaluation of Event Detection}



\begin{figure}
    \begin{minipage}{.22\textwidth}
    	\includegraphics[width=\linewidth]{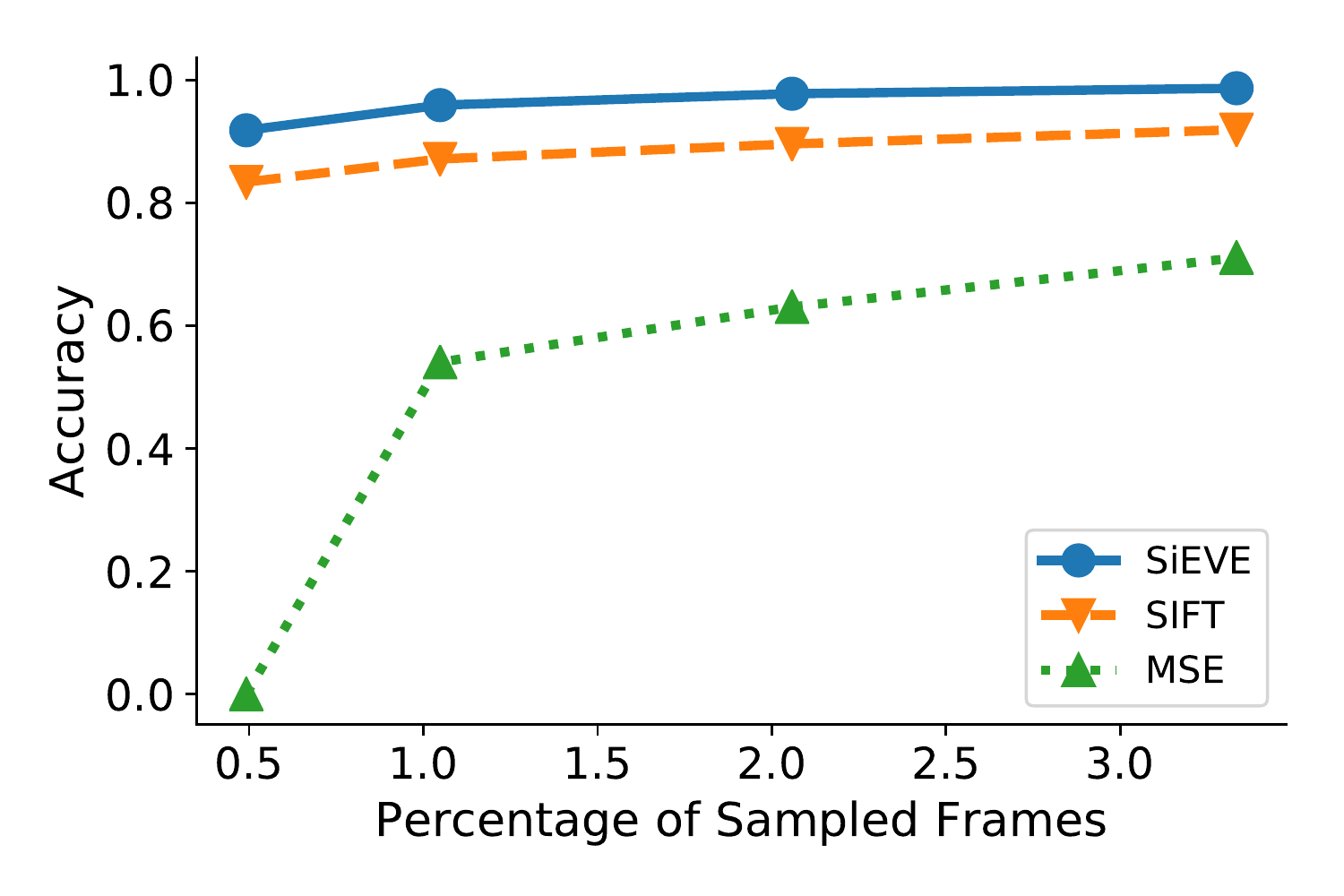}
    \end{minipage}
    \begin{minipage}{0.22\textwidth}
     	\includegraphics[width=\linewidth]{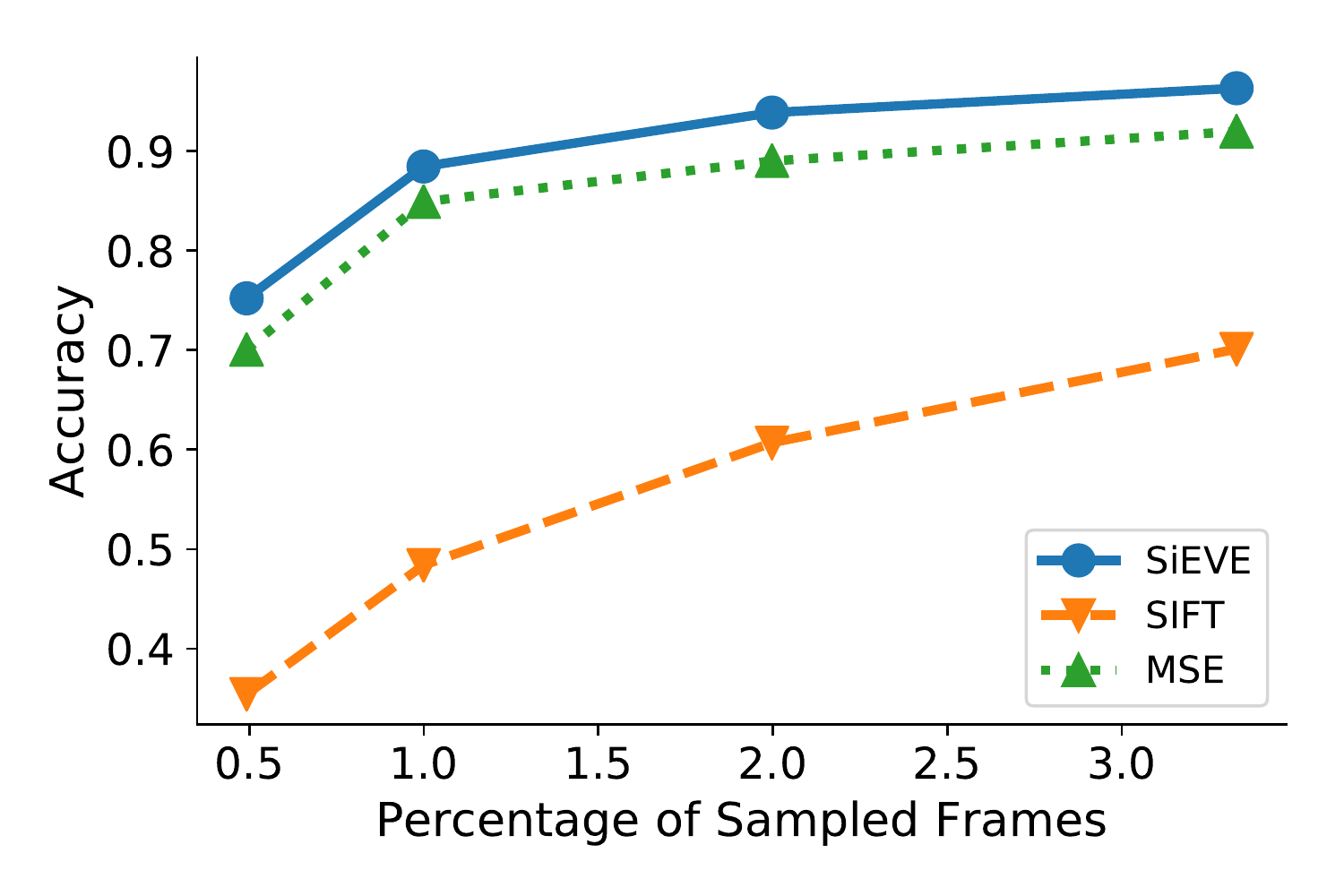}
    \end{minipage}
\caption{Accuracy at different sampling rates for Jackson sq. dataset (left) and Coral Reef dataset 2 (right)}
\label{fig:accuracy}
\vspace{-0.2in}
\end{figure}

{\bf Metrics:} In this section, we evaluate our method for event detection  (semantic video encoder + I-frame seeker) with other approaches. We use the following  metrics for comparison:

(1) {\bf Accuracy of per-frame object detection}: measured by percentage of frames with correct object labels with respect to the total number of frames.

(2) {\bf Percentage of sampled frames (SS)}: The percentage of frames that pass the I-frame seeker and undergo the NN processing with respect to the total number of video frames. 

(3) {\bf Speed of execution}: measured by the number of frames per seconds (fps) that can be processed by the event detection module. 

\noindent {\bf Compared Approaches:} we compare our approach with two other approaches. The two approaches rely on decoding each frame and computing an image similarity metric between the current frame and the previous frame. If the similarity is below a certain threshold, an event is detected. 
The frames before the next event are assigned the same object labels as the previous frame.
The approaches we evaluate are:

    (1) {\bf Mean squared error (MSE):} pixel-by-pixel mean squared difference between consecutive frames.
    
    (2) {\bf SIFT feature matching:} SIFT features are computed for each decoded frame and matched with the previous frame.
    

{\bf Accuracy:} We compare between the approaches based on the accuracy of per-frame object labels at different sampling rates. For example, we try different configurations of GOP size and scenecut threshold for SiEVE. Each of them gives a different number/position of I-frames and hence different accuracy. We show the accuracy of per-frame object detection when the number of I-frames is between 0.5\% and 3.5\% of the entire video stream. We tune the thresholds for other approaches to give the same sampling rate as SiEVE and we compare between the approaches in terms of accuracy at each sampling percentage. We present the results for the three datasets that we have ground truth labels for. We show the results for the first two datasets in Figure~\ref{fig:accuracy}. 
Due to space limitations, we omit the figure for the third dataset and we describe the summary of the results. The results show that for the three datasets SiEVE can achieve more than 95\% accuracy with analyzing 3.5\% of the video frames. SiEVE outperforms the related approaches by a significant margin in the three datasets. For the first dataset, SiEVE outperforms SIFT and MSE by an average of 11\% and 48\%, respectively. In the second dataset, SiEVE outperforms SIFT by 35\% and MSE by 8\%, and in the third dataset SiEVE outperforms SIFT by 28\% and MSE by 7\%. 
An interesting observation in the second and third datasets is that contrary to the first dataset, MSE outperforms SIFT. This is due to the different objects that are being detected in each dataset. MSE is well suited for detecting small objects (e.g., person, boat from long view) entering and leaving the scene which is the case in the second and third datasets. However, SIFT performs better for bigger objects (e.g., cars in close-up view) that cause significant changes in the scene. The proposed approach benefits from tuning the scenecut threshold to detect bigger or small objects. If a video has multiple labelled objects, the estimated  scenecut threshold tends to be tuned towards detecting the object that appears smaller in front of the camera. A smaller scene cut threshold is guaranteed to detect the existence of bigger objects as well because they create more motion.

{\bf Semantic Encoding vs Default Encoding Parameters:} We notice that the semantic encoding parameters that produce the best F1-score are different based on the nature of each video and are different from the default parameters (i.e., {\it sc=40, GOP-250}). This justifies why we tune the parameters separately for each camera feed. For the {\it sc} threshold, the tuned values are $100$, $200$, and $250$, for the first 3 videos where {\it sc} value of $250$ is more sensitive to small motion than $100$. The values are consistent with the relative sizes of the objects in front of the camera. For example, the first two videos have a close-up scene on the vehicles and people so they appear bigger in front of the camera and they create more motion compared to boats in the third video that was shot from a long distance. On the other hand, the $GOP$ sizes are $500$, $100$, $1000$, which are also related to how frequent objects appear in the video. For example, the people in the aquarium appear more frequently than the boats. We show a comparison between semantic encoding parameters in terms of accuracy, sample size (SS), and F1-score in Table~\ref{tab:params}. 

\begin{figure}
\centering
 \begin{adjustbox}{width=0.45\textwidth}
\begin{tabular}{|c|c|c|c|c|c|c|}
    \hline
    {\bf Dataset} & 
    \multicolumn{3}{|c|}{\bfseries Semantic}  &
    \multicolumn{3}{|c|}{\bfseries Default}\\ \cline{2-7}
    & Acc&SS&F1&Acc&SS&F1 \\ \hline
   
    Jackson sq. & 98.3\% & 2.1\% & 98.1\% & 72.6\% & 0.72\% & 83.9\%  \\ \hline
Coral reef & 99.1\% & 2.8\% & 98.16\% & 67.8\% & 0.75\% & 80.7\% \\ \hline
Venice & 96.5\% & 1.1\% & 97.6\% & 83.8\% & 0.4\% & 91\% \\ \hline

\end{tabular}
\end{adjustbox}
\captionof{table}{Comparison between semantic and default parameters in terms of accuracy (Acc), sample size (SS) and F1}
\label{tab:params}
\vspace{-0.1in}
\end{figure}

{\bf Speed of Execution:} The most significant improvement in our approach lies in the speed of performing event detection which is 100-124x faster than the closest image similarity approach. We show the results for the three datasets in Table~\ref{tab:speed}. SiEVE performs a lightweight computation in which it seeks the I-frames within a video which takes only 0.43 ms/frame (2300 fps) for 1080p frame resolution (dataset 3). On the other hand, the other approaches are bounded by time for decoding each video frame which takes 8 ms/frame (120 fps) for the same frame resolution. In addition to frame decoding, computing image features and image similarity drives the speed down to 22 fps for MSE and 16 fps for SIFT which results in 104x and 142x slowdown compared to SiEVE.
We notice that the same gain is carried over to the small resolution of 600x400 pixels. The speedup of SiEVE is 124x over MSE and 170x over SIFT.

  

  	


\begin{figure}
\centering
\begin{adjustbox}{width=0.3\textwidth}
\begin{tabular}{|c|c|c|c|}
\hline \bf Dataset  & \bf SiEVE & \bf MSE & \bf SIFT \\ \hline
Jackson square & 19600 & 157 & 115  \\ \hline
Coral reef & 7200 & 62 & 38 \\ \hline
Venice & 2300 & 22 & 16 \\ \hline
\end{tabular}
\end{adjustbox}
\captionof{table}{Speed of event detection results in terms of how many frames can be processed per second (fps) }
\label{tab:speed}
\vspace{-0.2in}

\end{figure}

\begin{figure}
 \centering
    \begin{minipage}{0.22\textwidth}
     \includegraphics[width=\linewidth]{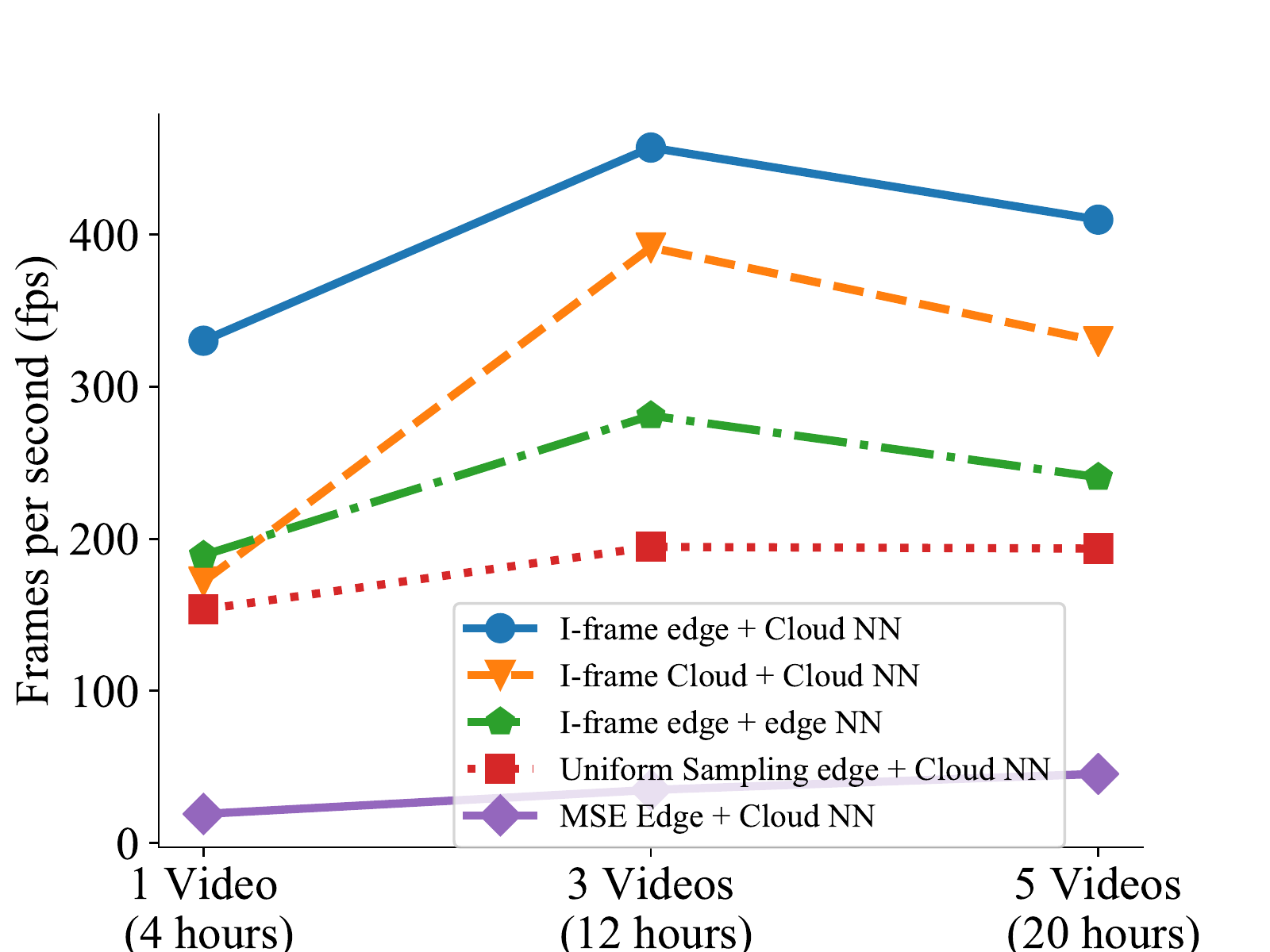}
	\caption{Number of processed frames per second by different baselines.}
	\label{fig:end-to-end}
    \end{minipage}
     \begin{minipage}{0.22\textwidth}
     \includegraphics[width=\linewidth]{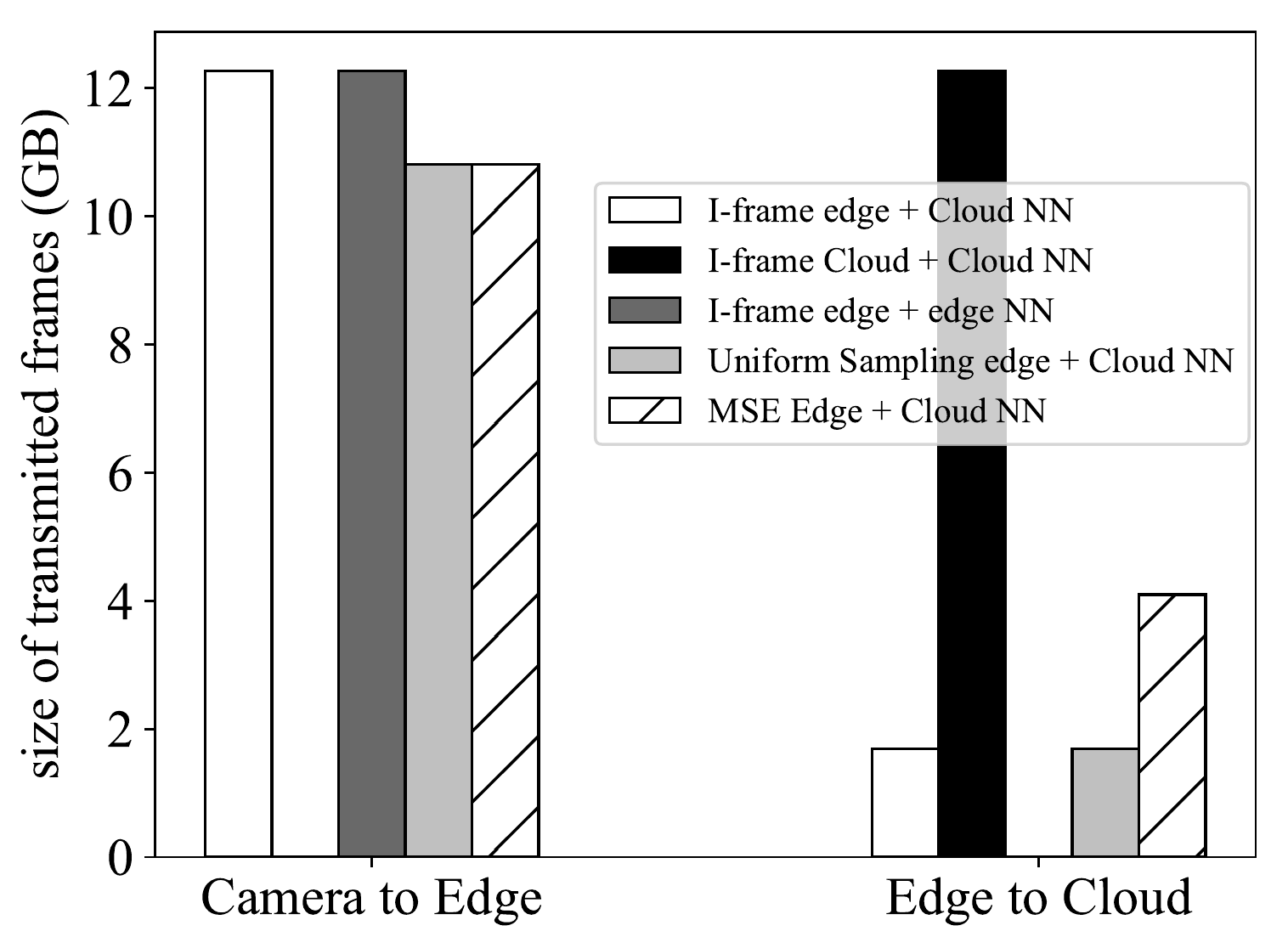}
	\caption{Total amount of data transfer for different baselines.}
	\label{fig:data-transfer}
    \end{minipage}

\end{figure}

\subsection{Evaluation of End-to-End System}\label{sec:end-to-end}
In this section, we evaluate the end-to-end performance of the system with respect to the system's throughput and the amount of data transmission from the edge to the cloud. We evaluate the throughput in the post-event analysis scenario in which the semantically encoded videos are pre-recorded and stored in the edge server and we use the five videos specified in Table~\ref{tab:datasets}. We use 4 hours from each video with the total of 20 hours. We compare our method with the following baselines:



(1) {\bf I-frame edge + cloud NN}: I-frame seeking in the edge, and  NN inference in the cloud. 
    
(2) {\bf I-frame edge + edge NN}: I-frame seeking in the edge and  NN inference in the edge. 
     
(3) {\bf I-frame cloud + cloud NN}: full video is streamed to the cloud where both I-frame seeking and NN inference are performed. 
    
(4) {\bf Uniform Sampling:} This approach includes uniformly sampling frames in the edge at fixed intervals, and transmitting the first frame in each interval for NN inference in the cloud. For fair comparison, we set the interval such that the number of transmitted frames is equal to the number of I-frames transmitted by the previous baselines. 
    
(5) {\bf MSE Edge + Cloud NN:} This approach includes executing MSE at the edge, and transmitting only the frames that pass a certain threshold to the cloud for inference.

    



The first three methods that implement I-frame seeking operate on semantically encoded videos while the other two methods operate on the video with the default encoding parameters. We chose the threshold of {\it MSE} and the semantic encoding parameters that achieves an F1-score of 95\% in the training set. For the two videos that we do not have ground truth labels we set the I-frame rate to 1 frame per 5 seconds for both approaches. The total number of frames in the 5 videos including I and P frames is 2.16 millions. Figure~\ref{fig:end-to-end} shows the throughput results in terms of the number of frames per seconds (fps) (i.e., $total\ number\ of\ frames$ / $total\ time\ in\ sec\ to\ process\ all\ frames$). From the results, we observe two important insights: (1) The first 3 methods that require semantic encoding significantly outperform the other two baselines including the lightweight uniform sampling. The reason is that uniform sampling requires decoding a large amount of P-frames unlike semantic encoding which focuses on I-frames only.  (2) We notice that the 3-tier architecture (i.e., camera, edge, cloud) leveraged in the first approach achieves significant speedup compared to the 2-tier architecture leveraged in the second two approaches (i.e., cloud only or edge only) because the 3-tier architecture benefits from the data filtering at the edge and the fast NN inference at the cloud.


We show the results for the amount of data filtering in the edge in Figure~\ref{fig:data-transfer}. The figure shows the amount of data transfer from camera to edge and from edge to cloud. We use YoloV3 as the NN inference model. From our experiments, we note that one of the limitations of semantically encoded videos is that they tend to have more I-frames than the original video. Hence, the data transmitted from the camera to the edge is 12\% larger than the original video.  However, after extracting I-frames and resizing them to the resolution of the YOLO model (i.e., 300x300), the size of the transmitted data from the edge to the cloud is reduced by a factor 7 (12.26GB to 1.688GB for I-frame edge + NN Cloud). Moreover, we note that the size of data transmitted by MSE is 2.5x larger than the aggregate size of I-frames which makes semantic encoding and I-frame seeking a more bandwidth-efficient approach.


\section{Acknowledgements}\label{sec:ack}

\noindent This research was conducted during summer internship of Tarek Elgamal in AT\&T Labs and the publication efforts were supported by the Ralph and Catherine Fisher Professorship funds.

\section{Conclusion}\label{sec:conc}


In this paper, we present SiEVE, a 3-tier video analytics system to reduce the latency and increase the throughput of NN-based analysis over video streams. In SiEVE, we address the problem of semantic video encoding in which the video encoder becomes aware of the downstream object detection task. We show that video encoders can produce I-frames when an object enters or leaves the scenes. This allows the video to be analyzed through seeking I-frames only rather than decoding the entire video which results in 100x speedup compared to decoding each frame and running image similarity.




\end{document}